\def\year{2022}\relax
\documentclass[letterpaper]{article} % DO NOT CHANGE THIS
\usepackage{aaai22}  % DO NOT CHANGE THIS
\usepackage{times}  % DO NOT CHANGE THIS
\usepackage{helvet}  % DO NOT CHANGE THIS
\usepackage{courier}  % DO NOT CHANGE THIS
\usepackage[hyphens]{url}  % DO NOT CHANGE THIS
\usepackage{graphicx} % DO NOT CHANGE THIS
\usepackage{natbib}  % DO NOT CHANGE THIS AND DO NOT ADD ANY OPTIONS TO IT
\usepackage{caption} % DO NOT CHANGE THIS AND DO NOT ADD ANY OPTIONS TO IT
\DeclareCaptionStyle{ruled}{labelfont=normalfont,labelsep=colon,strut=off} % DO NOT CHANGE THIS
\usepackage{algorithm}
\usepackage{algorithmic}
\usepackage{xcolor}
\usepackage{lscape}

\usepackage{array}
\newcolumntype{P}[1]{>{\raggedright\arraybackslash}p{#1}}
\usepackage{newfloat}
\usepackage{listings}
\floatstyle{ruled}
\newfloat{listing}{tb}{lst}{}
\floatname{listing}{Listing}
\title{BillionCOV: An Enriched Billion-scale Collection of COVID-19 tweets for Efficient Hydration}

\author{
    Rabindra Lamsal, Maria Rodriguez Read, Shanika Karunasekera
}
\affiliations{
    School of Computing and Information Systems\\
    The University of Melbourne\\
    Melbourne, Victoria 3010\\
    \{r.lamsal,maria.read,karus\}@unimelb.edu.au
}

\usepackage{bibentry}
\begin{document}
\nocopyright
\maketitle

\begin{abstract}
The COVID-19 pandemic introduced new norms such as social distancing, face masks, quarantine, lockdowns, travel restrictions, work/study from home, and business closures, to name a few. The pandemic's seriousness made people vocal on social media, especially on microblogs such as Twitter. Researchers have been collecting and sharing large-scale datasets of COVID-19 tweets since the early days of the outbreak. Sharing raw Twitter data with third parties is restricted; users need to hydrate tweet identifiers in a public dataset to re-create the dataset locally. Large-scale datasets that include original tweets, retweets, quotes, and replies have tweets in billions which takes months to hydrate. The existing datasets carry issues related to \textit{proportion} and \textit{redundancy}. We report that more than 500 million tweet identifiers point to deleted or protected tweets. In order to address these issues, this paper introduces an enriched global billion-scale English-language COVID-19 tweets dataset, \textbf{BillionCOV}\footnote{\texttt{https://dx.doi.org/10.21227/871g-yp65}}, that contains 1.4 billion tweets originating from 240 countries and territories between October 2019 and April 2022. Importantly, \textit{BillionCOV} facilitates researchers to filter tweet identifiers for efficient hydration. This paper discusses associated methods to fetch raw Twitter data for a set of tweet identifiers, presents multiple tweets' distributions to provide an overview of \textit{BillionCOV}, and finally, reviews the dataset's potential use cases.
\end{abstract}

\section{Introduction}
As of December 20, 2022, the confirmed cases of Coronavirus disease 2019 (COVID-19) have reached over 658 million, with 6.67 million deaths and 632 million recovered cases \cite{worldometer2020coronavirus}. The virus was first identified in an outbreak in Wuhan, China, in December 2019, and shortly after a few weeks, it spread to other regions of China and subsequently worldwide. The World Health Organization (WHO) declared the outbreak a public health emergency of international concern on January 30, 2020, and a pandemic on March 11, 2020. States and territories worldwide attempted to contain the spread of the virus by initiating strict lockdowns and even curfews. Since the outbreak, global citizens have had to adjust their lifestyles with new norms such as social distancing, face masks, quarantine,  lockdowns, travel restrictions, and business closures. The seriousness of the pandemic made people significantly verbal on social media, particularly on microblog platforms such as Twitter.

Twitter discussions, i.e., tweets, regarding the pandemic have been reported to be in hundreds of millions. Numerous COVID-19 tweet collections \cite{chen2020tracking,lamsal2021design,banda2021large,imran2022tbcov} have been released, anticipating that they would assist researchers in the crisis informatics domain to explore the conversational dynamics of the pandemic through diverse sets of spatial and temporal analyses. Such datasets release tweet identifiers to comply with Twitter's data redistribution policy\footnote{https://developer.twitter.com/en/developer-terms/policy/}. Each tweet has a unique identifier on the platform. The identifiers in a public dataset should be ``hydrated" to re-create it locally. Fetching raw Twitter data for a set of identifiers from Twitter servers using their tweet lookup endpoint is known as ``hydration of tweet identifiers". To maintain reliability and scalability, Twitter places rate limits on the number of requests that can be made to its APIs. Each of their endpoints has different rate limits\footnote{https://developer.twitter.com/en/docs/twitter-api/rate-limits/}. For instance, their tweet lookup endpoint, which is used for hydrating tweet identifiers, allows up to 900 requests per 15-minute window. With each request fetching 100 tweets, 8.64 million tweet identifiers can be hydrated in a day. 

However, tweets in \cite{chen2020tracking,lamsal2021design,imran2022tbcov} are above 2 billion, and at a rate of 8.6 million tweets per day, it would take more than seven months to hydrate one of these datasets. COVID-19 datasets sharing geotagged tweets and region-specific tweets are comparatively small and can be hydrated within a reasonable time; however, global datasets that include original tweets, retweets, quotes, and replies have tweets in billions which takes months to hydrate. Also, the majority of the existing large-scale datasets are multilingual, thus reporting fewer English-language tweets because of limits associated with Twitter's filtered stream endpoint. Furthermore, some existing datasets provide none to a few sets of additional tweet objects for filtering tweet identifiers, and some have limited temporal coverage. Therefore, in order to address these issues, we introduce an enriched global billion-scale English-language COVID-19 tweets dataset, \textit{BillionCOV}, which facilitates researchers to filter tweet identifiers before hydration. Also, this paper serves as a tutorial on tweet hydration as we discuss associated methods to fetch raw Twitter data for a set of tweet identifiers.

\section{Related Work}
Researchers and laboratories worldwide have been collecting and sharing multiple COVID-19 tweets datasets. Some of them are multilingual \cite{chen2020tracking,banda2021large,imran2022tbcov}, while some are language-specific \cite{alqurashi2020large,haouari2020arcov,lamsalIEEEa,lamsalIEEEb} and region-specific \cite{lamsal2022twitter}. Hydrating existing global COVID-19 datasets can take a significant amount of time, as they usually have from hundreds of millions to billions of tweets. \cite{chen2020tracking} maintain a multilingual COVID-19 tweets repository and release only tweet identifiers. The oldest tweets in their dataset date back to January 21, 2020, and as per their latest release (v2.103), the dataset has 2.678 billion tweets. Similarly, \cite{banda2021large} maintain a multilingual repository with the oldest tweets dating back to January 1, 2020, and share tweet identifiers and date/time information, and additional tweet objects (for tweets after August 2020)~---~language and country code. \cite{lamsalIEEEa} maintains a repository of more than 2.2 billion tweet identifiers and their respective sentiment scores. The oldest tweets in the dataset date back to October 1, 2019. \cite{imran2022tbcov} share 2 billion tweet identifiers alongside tweet objects such as date/time, language, user identifier, replies/quotes labels, sentiment scores, and geo-information. The oldest tweets in their dataset date back to February 1, 2020.

\subsection{Issue with existing datasets}
Billion-scale multilingual datasets raise issues related to \textit{proportion} and \textit{redundancy}. The \textit{proportion} issue exists due to limits placed by Twitter on its filtered stream endpoint. Since 450 requests are allowed per 15-minute window per application, a maximum of 4.32 million tweets can be fetched in 24 hours. Also, the endpoint's payload returns 1\% of the entire Twitter data at a particular time. The language-based distribution of tweets in multilingual datasets shows a higher prevalence of English, Spanish, Portuguese, French, and Indonesian languages \cite{chen2020tracking,banda2021large,imran2022tbcov}. As a result, multi-lingual datasets contain fewer tweets for a language unless numerous language-dedicated data collections are done and merged later. For instance, \cite{chen2020tracking} report the presence of 1.7 billion English-language tweets in their multi-lingual corpus of 2.678 billion tweets, while \cite{lamsalIEEEa} reports 2.2 billion English-language tweets in their English-only collection. Regarding the \textit{redundancy} issue, the size and/or multi-lingual nature of the existing datasets become a concern for researchers who want to hydrate only English-language tweets, geo-specific, or certain tweet types (original tweets, retweets, quotes, replies). The entire dataset needs hydration and later filtration to re-create the desired dataset. \cite{banda2021large,imran2022tbcov} do provide additional tweet objects to help filter tweet identifiers before hydration; however, \cite{banda2021large} provide a limited set of tweet objects and \cite{imran2022tbcov}, with a comprehensive list of tweet objects, received its last release on March 31, 2021. Tweet identifiers can also point to either deleted or protected tweets. None of the datasets in the literature seem to filter out those kinds of tweets.

As a contribution to the literature, we introduce an enriched global billion-scale English-language COVID-19 tweets dataset, \textit{BillionCOV}, which solves the issues related to \textit{proportion} and \textit{redundancy}. \textit{BillionCOV} is an English-only collection and therefore addresses the \textit{proportion} issue for English language only. And regarding the \textit{redundancy} issue, \textit{BillionCOV} is curated by filtering out deleted and protected tweets from \cite{lamsalIEEEa}, and most importantly, the dataset includes additional tweet data useful for filtering tweet identifiers before hydration. The dataset facilitates the filtration of tweet identifiers as per the following contexts: Is this a reply tweet? (TRUE/FALSE), Is this a retweet? (TRUE/FALSE), Is this a quote tweet? (TRUE/FALSE), Is the author of the tweet verified? (TRUE/FALSE), and country (e.g., US, AU, etc.).

\section{Data Curation}

\begin{figure}
    \centering
    \includegraphics[width=0.5\textwidth]{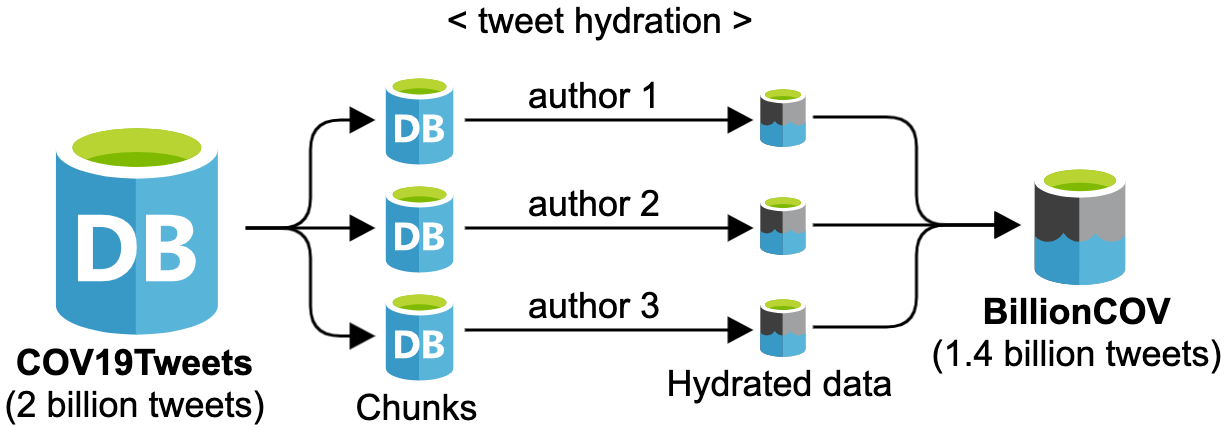}
    \caption{The data curation process. The three chunks of the \textit{COV19Tweets} dataset were hydrated separately by the three respective authors.}
    \label{twarc-hydration}
\end{figure}

We hydrated 2 billion tweet identifiers present in \textit{COV19Tweets} \cite{lamsalIEEEa}, following the process shown in Figure \ref{twarc-hydration}, using \textit{twarc} python library. Refer to \cite{lamsal2021design} for details regarding keywords and hashtags, endpoints, collection strategy, and infrastructure used for curating \textit{COV19Tweets}. At the time of hydration, the dataset contained tweets created between October 1, 2019, and April 27, 2022. With three authors hydrating the three separate chunks, it took us just over 2.5 months to fully hydrate 2 billion tweet identifiers. The hydration ran from August 2022 to October 2022. Out of 2 billion identifiers, we retrieved 1.4 billion tweets, with more than 500 million tweet identifiers pointing to either deleted or protected tweets. As per official reports\footnote{https://transparency.twitter.com/en/reports/covid19.html}, while enforcing the COVID-19 misleading information policy, Twitter challenged 11.72 million accounts, suspended 11,230 accounts, and removed over 97,674 content (worldwide) between January 2020 and September 2022. When accounts are suspended/removed, the retweets associated with those accounts are unavailable, and \cite{lamsalIEEEa} includes all forms of tweets, including retweets. Users may also delete or make their profile protected. All these factors resulted in the unavailability of around 25\% of the 2 billion tweets. Avoiding the hydration of solely the unavailable tweets saves almost two months in a single hydration task. Twitter provides a \textit{batch compliance} endpoint to check the availability of each tweet identifier (e.g., deleted, protected); however, \textit{BillionCOV} waives this extra effort for those 500 million tweet identifiers. Note that \textit{BillionCOV} might still contain tweet identifiers that point to deleted or protected tweets when hydrated since a tweet can be deleted or made protected by its author at any time.

There are numerous tools available for hydrating tweet identifiers. Some of the widely used ones are \textit{Hydrator} and \textit{twarc}. Hydrator\footnote{https://github.com/DocNow/hydrator/releases} is a desktop application, while twarc\footnote{https://twarc-project.readthedocs.io/en/latest/} is both a command line tool and a Python library. Archiving billions of tweets as JSONL data can take terabytes of space, so every use case might not require archiving the complete set of tweet objects. Some of the retrievable tweet objects are given in Table \ref{objects-list}. twarc provides more flexibility as we get to decide which tweet objects are to be retrieved. Both of these tools handle the rate limits set by Twitter across its endpoints. In this study, we used twarc and retrieved all possible tweet objects for each tweet identifier in \cite{lamsalIEEEa}.

\begin{table*}[ht]
    \centering
    \caption{Below are some of the retrievable tweet objects. For a description of each object, refer to the tweet lookup endpoint's documentation$^\#$.}
    \label{objects-list}
    \begin{tabular}{|P{17.3cm}|}
\hline
\textbf{General}: id, conversation id, creation datetime, tweet, language, source, reply settings, possibly sensitive, author id, in reply to user id, retweeted user id, quoted user id\\
\hline
\textbf{Referenced tweets}: replied to id, retweeted id, quoted id\\
\hline
\textbf{Public metrics}: like count, quote count, reply count, retweet count\\
\hline
\textbf{Edit controls}: edits remaining, editable until, is edit eligible\\
\hline
\textbf{Withheld}: scope, copyright, country codes \\ 
\hline
\textbf{Entities}: annotations, cashtags, hashtags, mentions, urls, contexts \\
\hline
\textbf{Attachments}: media, media keys, poll duration minutes, poll end datetime, poll id, poll options, poll voting status \\
\hline
\textbf{Author}: id, creation datetime, username, name, description, location, pinned tweet id, profile image url, protected, url, verified, withheld scope, withheld copyright, withheld country codes \\
\hline
\textbf{Author Entities}: description cashtags, description hashtags, description mentions, description urls \\
\hline
\textbf{Author public metrics}: followers count, following count, listed count, tweet count \\
\hline
\textbf{Geo}: coordinates, coordinates type, country, country code, full name, bounding box, name, place id, place type \\
\hline
    \end{tabular}
\raggedright $^\#$\url{https://developer.twitter.com/en/docs/twitter-api/tweets/lookup/api-reference/get-tweets}
\end{table*}

\section{Dataset Structure and Hydration}
\textit{BillionCOV} has 1.4 billion tweets, and if the use case requires hydrating all available tweets, note that it will take more than 5 months to re-create the dataset locally in a single hydration task. Running multiple such tasks through collaborations, without violating Twitter's terms of service, can decrease the hydration time significantly. We split the whole dataset into multiple parts, each with 50 million tweets. Consider concatenating or further splitting the files depending on the machine's memory where hydration is planned. Once the hydration completes, the JSONL data can be exported to CSV\footnote{https://github.com/DocNow/twarc-csv} and used as \textit{pandas} DataFrame for analysis. Parallel computing libraries such as \textit{Dask}\footnote{https://docs.dask.org/en/stable/} can be used if the resulting CSV is not loadable into the memory with \textit{pandas}.

However, if the use case needs a specific set of tweets, the tweet identifiers need filtration. The additional tweet data we provide in \textit{BillionCOV} are as follows: 

\begin{itemize}
    \item (\textit{id}) refers to the unique identifier of a tweet
    \item (\textit{is\_reply}) refers to ``Is this a reply tweet?"
    \item (\textit{is\_retweet}) refers to ``Is this a retweet?"
    \item (\textit{is\_quote}) refers to ``Is this a quote tweet?"
    \item (\textit{country}) refers to country code
\end{itemize}

After hydration, we obtained over 90 tweet objects, some of which are listed in Table \ref{objects-list}. The reference tweet metadata in the hydrated dataset included fields for \texttt{replied to id}, \texttt{retweeted id}, and \texttt{quoted id}. These fields referred to the identifier of the referenced tweet based on whether a tweet was a reply, retweet, or quoted tweet. We used this information to generate data for \texttt{is\_reply}, \texttt{is\_retweet}, and \texttt{is\_quote}, with these three columns containing boolean values of TRUE or FALSE. If a tweet in the dataset has all three of these values as FALSE, it is considered to be an original tweet.

We do not release tweet creation date/time as this metadata is redundant. Twitter generates tweet identifiers based on timestamps, which are not sequential. Algorithms \ref{timestamp-1} and \ref{timestamp-2} show the Python implementation for converting a tweet identifier to its human-readable timestamp and vice versa. Using this information, the tweet identifiers can be filtered at the temporal level and then hydrated.

In either use case, hydrating complete or specific set of \textit{BillionCOV}, there are two ways~---~archiving all tweet objects or only the selected ones. Algorithm \ref{twarc-1} presents a command line use of twarc for archiving all tweet objects, and Algorithm \ref{twarc-2} explains the usage of twarc as a python library for archiving a selected set of tweet objects. Refer to Appendix A for \textit{tweet data dictionary}, which provides a comprehensive list of tweet objects retrievable using twarc.

\begin{algorithm}
\caption{Conversion of tweet identifier to timestamp}
\label{timestamp-1}
\textbf{Input}: tweet identifier\\
\textbf{Output}: timestamp
\begin{algorithmic}[1]
\STATE import time
\STATE tweet\_id=\textit{tweet~id~here}

\STATE shifted\_id=tweet\_id $>>$ 22 \COMMENT{applying right shift operator on tweet ID}
\STATE timestamp=shifted\_id + 1288834974657 \COMMENT{the default Twitter epoch is equivalent to November 04, 2010 01:42:55 AM UTC}
\STATE data\_time=time.ctime(timestamp/1000)

\end{algorithmic}
\end{algorithm}

\begin{algorithm}
\caption{Conversion of timestamp to tweet identifier}
\label{timestamp-2}
\textbf{Input}: timestamp\\
\textbf{Output}: tweet identifier
\begin{algorithmic}[1]

\STATE import datetime
\STATE epoch=datetime.datetime.utcfromtimestamp(0)
\STATE dt=datetime.datetime(\textit{timestamp~here}) \COMMENT{timestamp format: year, month, day, hour, minute, second, microsecond; e.g., 2022, 1, 1, 0, 0, 0, 000000}
\STATE milisecond\_epoch=int((dt-epoch).total\_seconds()*1000)
\STATE epoch=milisecond\_epoch - 1288834974657
\STATE tweet\_id=epoch $<<$ 22 \COMMENT{applying left shift operator}

\end{algorithmic}
\end{algorithm}

\begin{algorithm}
\caption{Archiving all tweet objects using twarc's command line functionality}
\label{twarc-1}
\textbf{Input}: txt/csv file containing tweet identifiers on each line, without quotes or header \\
\textbf{Output}: hydrated JSONL data
\begin{algorithmic}[1]
\vspace{0.2cm}

\STATE twarc2 hydrate {-}{-}consumer-key \textit{your-consumer-key-here} {-}{-}consumer-secret \textit{your-consumer-secret-here} {-}{-}access-token \textit{your-access-token-here} {-}{-}access-token-secret \textit{your-access-token-secret-here} \textit{your-ids-file.txt} \textit{where/to/save/the/hydrated/data.jsonl}
\vspace{0.2cm}

\STATE nohup \textit{hydration-command-here} $>$ output.out 2$>$\&1 \& \COMMENT{Hydration is a time-consuming task. Use \textit{nohup} to continue the task in the background. The progress output is saved in output.out.}

\end{algorithmic}
\end{algorithm}

\begin{algorithm}[ht!]
\caption{Archiving selected tweet objects with twarc}
\label{twarc-2}
\textbf{Input}: txt/csv file containing tweet identifiers on each line, without quotes or header \\
\textbf{Output}: hydrated JSONL data
\begin{algorithmic}[1]
\STATE from twarc import Twarc

\STATE consumer\_key=``\textit{consumer-key-here}"
\STATE consumer\_secret=``\textit{consumer-secret-here}"
\STATE access\_token=``\textit{access-token-here}"
\STATE access\_token\_secret=``\textit{access-token-secret-here}"

\STATE t=Twarc(consumer\_key, consumer\_secret, access\_token, access\_token\_secret)

\FOR{tweet \textbf{in} t.hydrate(open(``\textbf{ids.txt}"))}
\STATE tweetID=tweet[``id\_str"]
\STATE tweetText=tweet[``full\_text"]
\STATE language=tweet[``lang"]
\STATE source=tweet[``source"]
\STATE repliedToId=tweet[``in\_reply\_to\_status\_id"] \COMMENT{returns referenced tweet's identifier, else \textit{None} if tweet is not a reply. The same notion applies to retweets and quoted tweets}
\STATE geoCoordinates=tweet[``coordinates"][``coordinates"] \COMMENT{Gives [lon,lat] pair if the tweet is geotagged with point coordinates}
\STATE country=tweet[``place"][``country"] \COMMENT{valid, if place information is available}

\STATE userCreated=tweet[``user"][``created\_at"]
\STATE userProfileLocation=tweet[``user"][``location"]
\STATE userFollowerCount=tweet[``user"][``followers\_count"]

\COMMENT{Similar procedure can be followed for the remaining objects. Refer to Appendix A for a detailed schema of the \textit{tweet data dictionary}.}
\ENDFOR

\end{algorithmic}
\end{algorithm}

\section{The BillionCOV Dataset}
In this section, we briefly explore the dataset, discuss its potential use cases and provide additional information.

\subsection{Description}

\begin{figure}
    \centering
\includegraphics[width=0.5\textwidth]{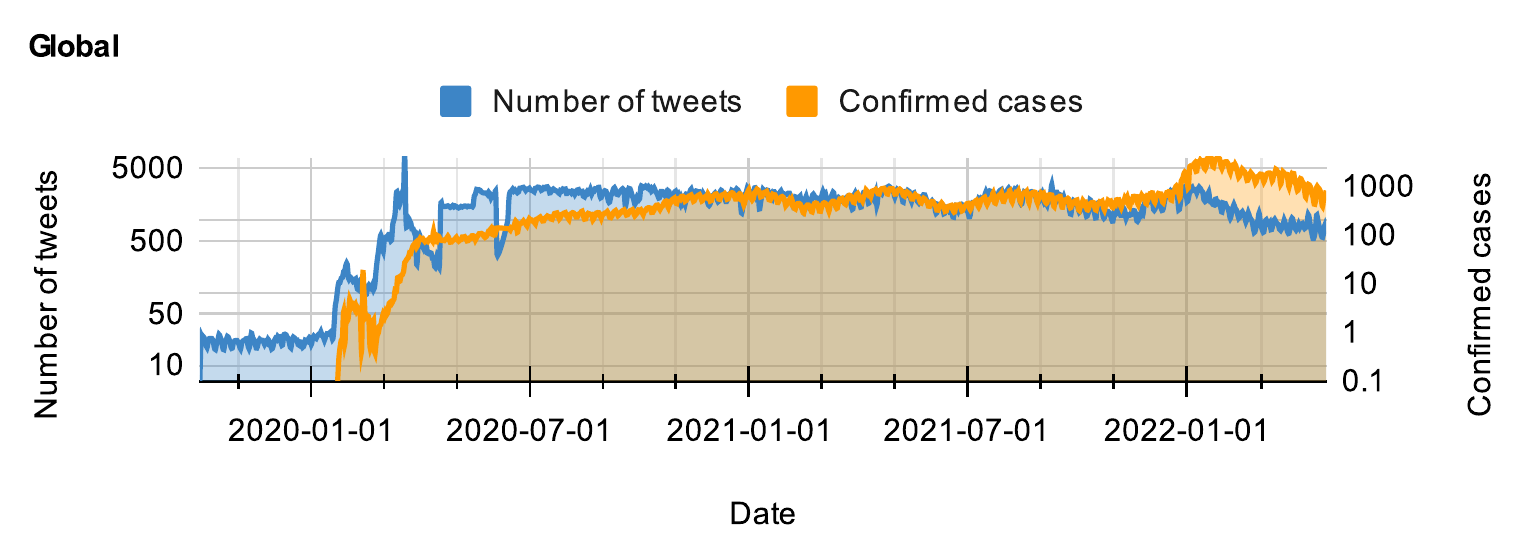}
\includegraphics[width=0.5\textwidth]{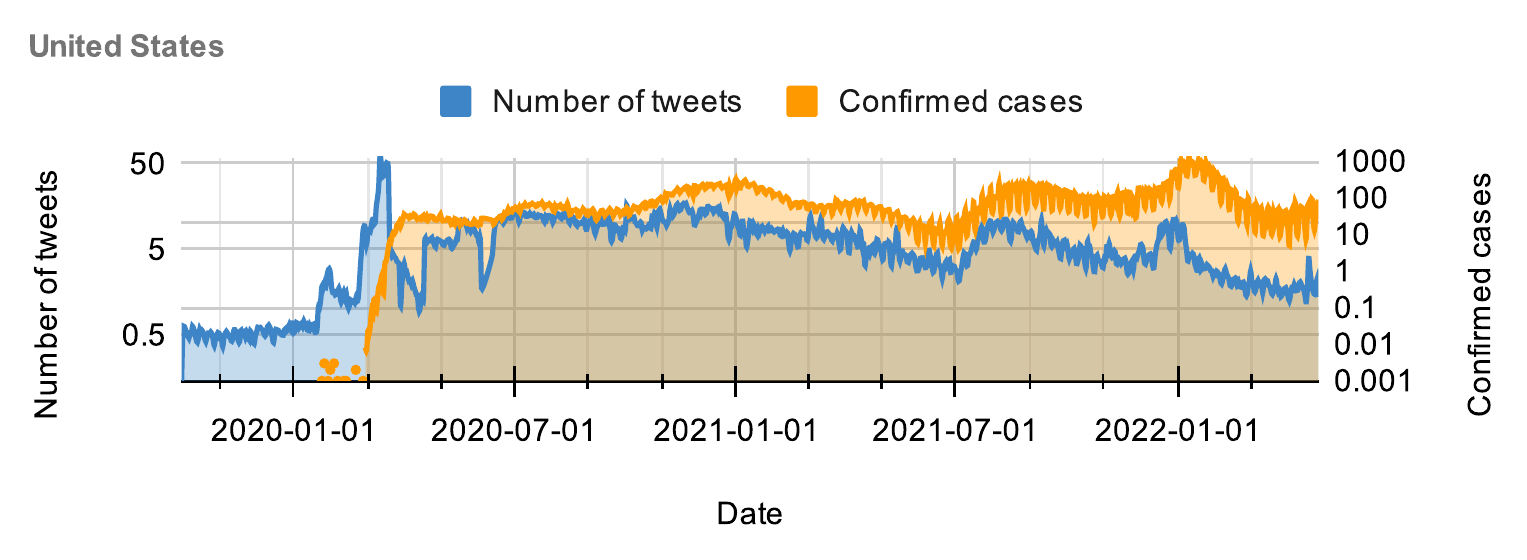}
\includegraphics[width=0.5\textwidth]{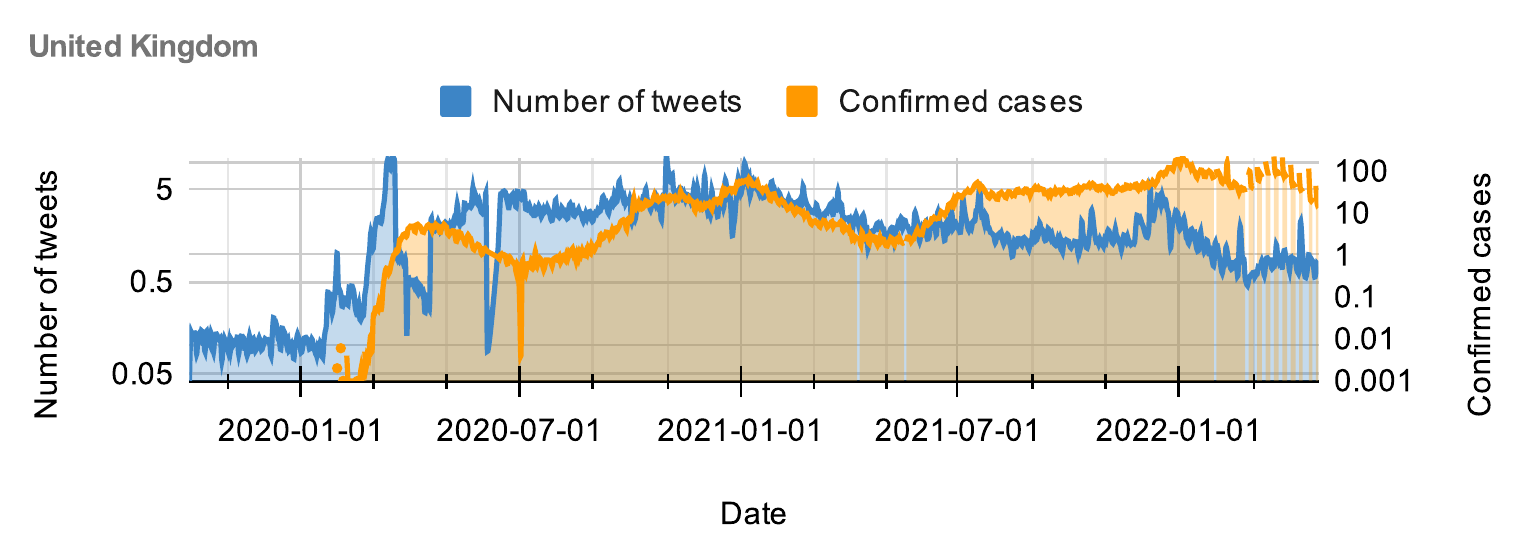}
\includegraphics[width=0.5\textwidth]{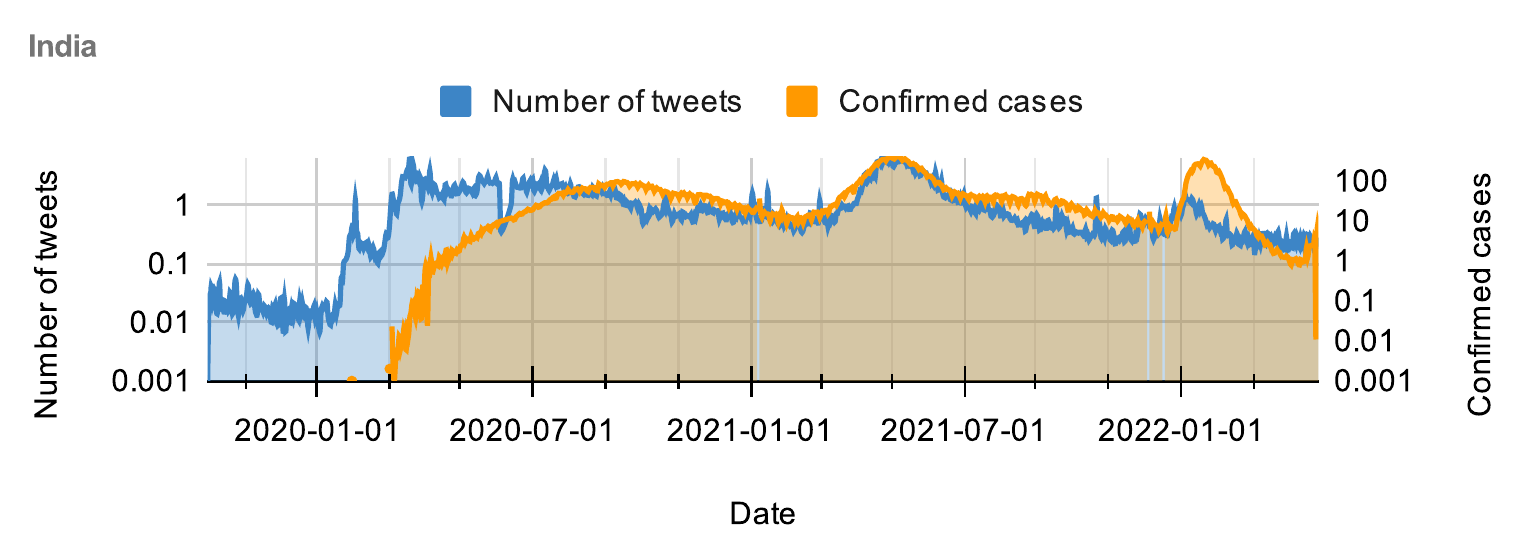}
\includegraphics[width=0.5\textwidth]{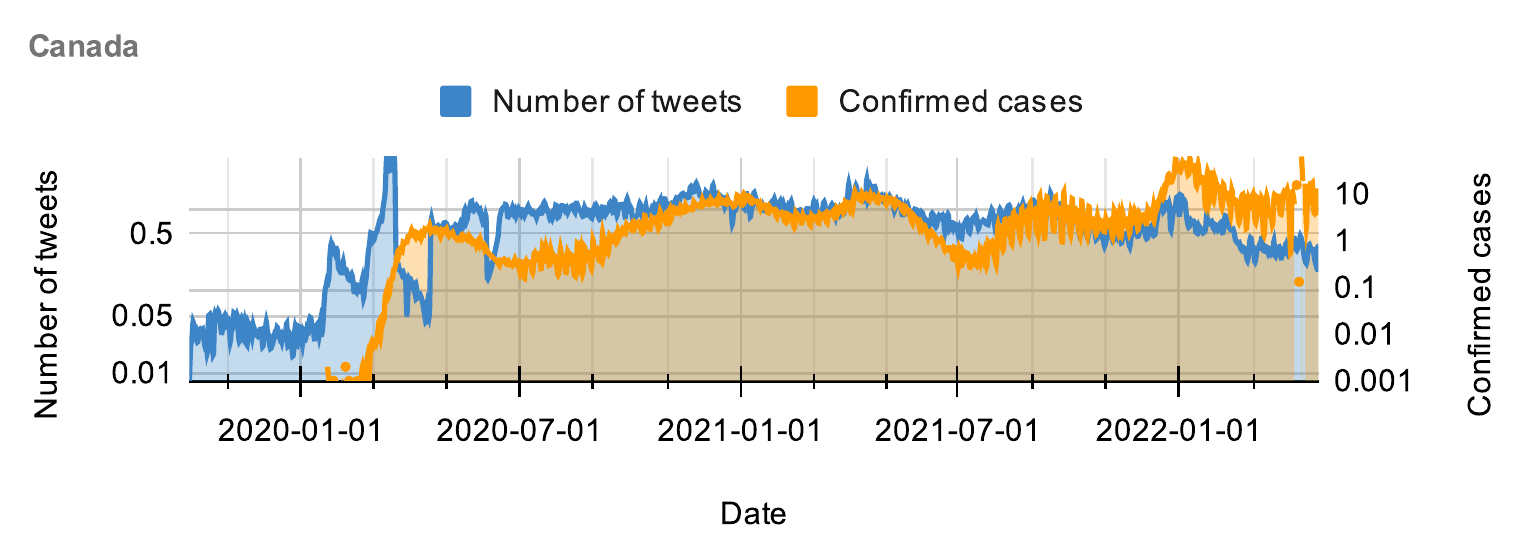}
\includegraphics[width=0.5\textwidth]{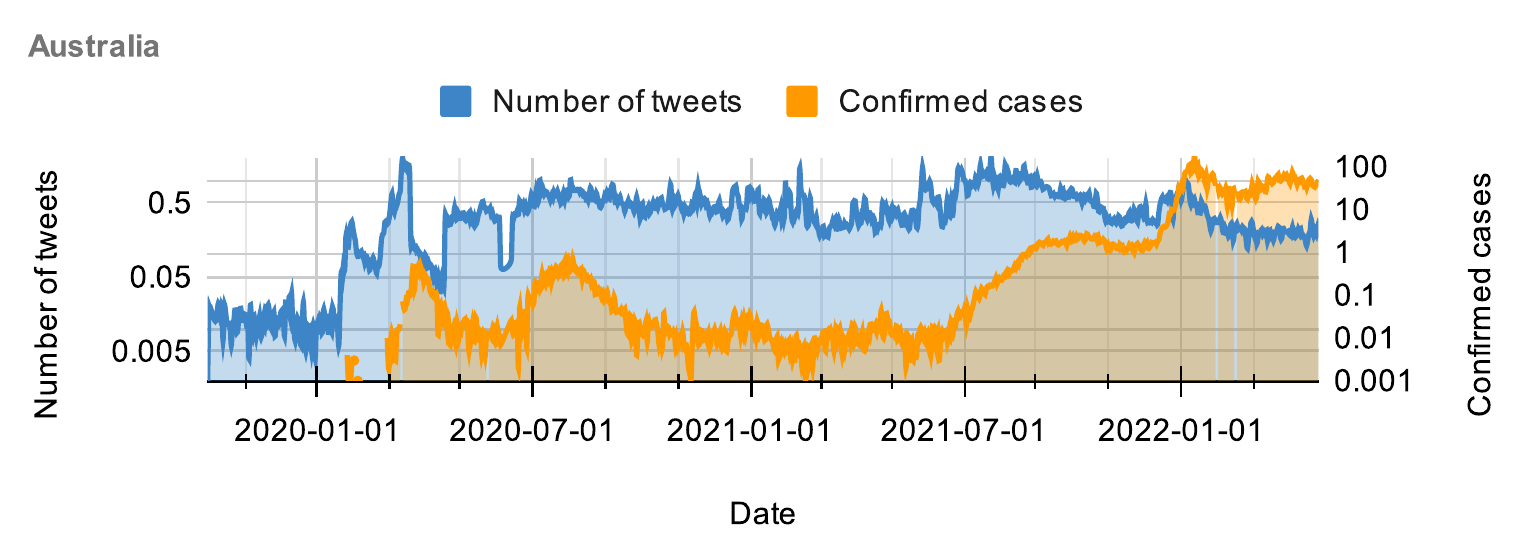}
\caption{Daily distributions of tweets and confirmed cases for the globe and top 5 countries in the discourse. Global distribution includes all tweets and country-specific distribution includes geotagged tweets. For all distributions, $Y$-axes are in log scale with a scale factor of 1000.}
\label{tweets-distributions}
\end{figure}

\textit{BillionCOV} has 1,410,446,121 English-language tweets regarding the COVID-19 pandemic, originating from 240 countries and territories between October 2019 and April 2022. \cite{lamsalIEEEa} used the Full-archive search endpoint to collect historical tweets beyond March 2020. The daily distributions of all tweets (for the globe) and geotagged tweets (for selected countries) alongside confirmed COVID-19 cases are presented in Figure \ref{tweets-distributions}. The data source for the confirmed cases is \textit{Our World in Data}\footnote{https://ourworldindata.org/covid-cases}. United States, United Kingdom, India, Canada, and Australia are the top 5 countries in the discourse (based on the \textit{geo.country} tweet object) and are followed by South Africa, Ireland, Nigeria, Philippines, and Malaysia in the top 10. A comprehensive list of countries participating in the discourse is in Table \ref{countries-list}.

\begin{figure}
    \centering
    \includegraphics[width=0.35\textwidth]{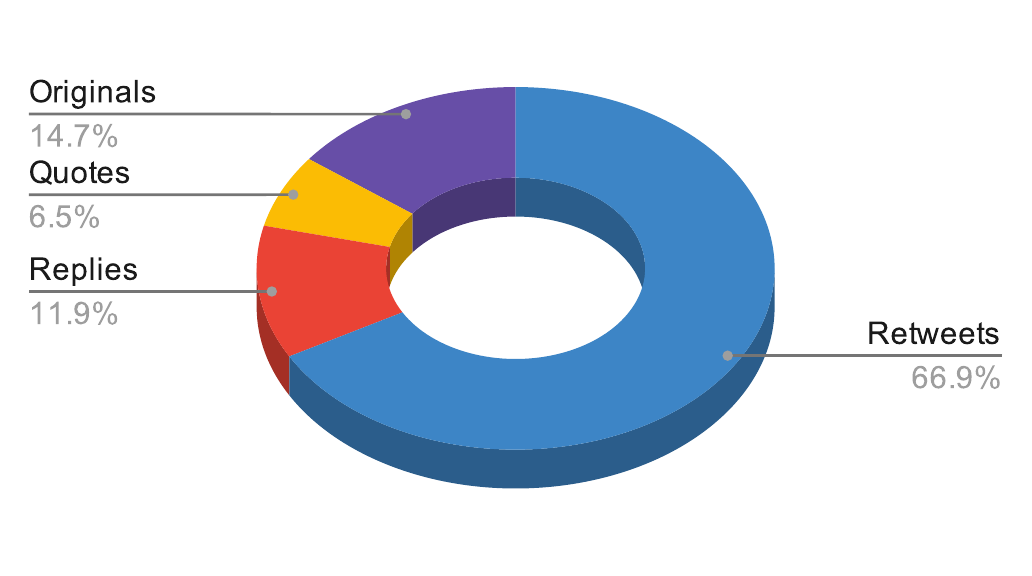}
    \caption{Proportion of original tweets, retweets, quote tweets, and reply tweets.}
    \label{tweet-types}
\end{figure}

\begin{figure}
    \centering
    \includegraphics[width=0.35\textwidth]{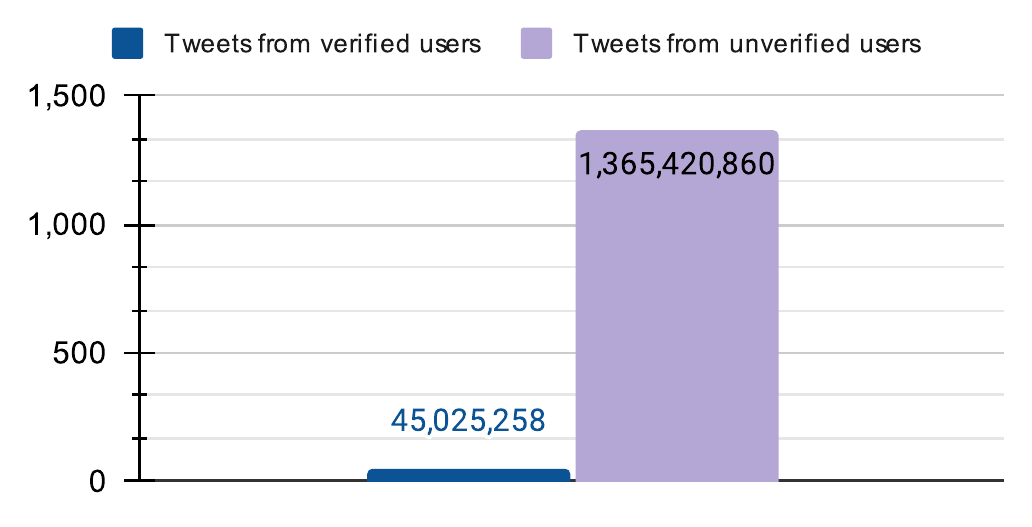}
    \caption{Tweets from verified users versus unverified users (scale factor of 1 million).}
    \label{verified-stats}
\end{figure}

\begin{table}
\centering
\caption{Top 100 countries in the discourse with respect to geotagged tweets. Table lists ISO alpha-2 country codes$^\#$.}
\label{countries-list}
\begin{tabular}{|>{\centering}m{0.4cm}|c|c|>{\centering}m{0.4cm}|c|c|}
\hline
\textbf{SN} & \textbf{Country} & \textbf{Tweets} & \textbf{SN} & \textbf{Country} & \textbf{Tweets}\\
\hline
1  & US & 5,982,937 & 51  & NO & 8,324 \\
2  & GB & 2,021,222 & 52  & LB & 8,231 \\
3  & IN & 1,011,205 & 53  & AR & 8,133 \\
4  & CA & 693,421  & 54  & PL & 8,020 \\
5  & AU & 393,476  & 55  & MV & 7,913 \\
6  & ZA & 267,037  & 56  & GR & 7,832 \\
7  & IE & 199,574  & 57  & AT & 7,525 \\
8  & NG & 167,662  & 58  & FI & 6,911 \\
9  & PH & 111,392  & 59  & FJ & 6,894 \\
10 & MY & 79,916   & 60  & BB & 6,766 \\
11 & KE & 74,437   & 61  & VN & 6,229 \\
12 & PK & 73,335   & 62  & MM & 6,205 \\
13 & NZ & 66,318   & 63  & TZ & 6,128 \\
14 & DE & 53,905   & 64  & YE & 6,018 \\
15 & GH & 47,040   & 65  & MW & 5,897 \\
16 & UG & 45,177   & 66  & CY & 5,284 \\
17 & ES & 44,971   & 67  & EG & 5,164 \\
18 & NL & 41,300   & 68  & OM & 4,762 \\
19 & FR & 38,605   & 69  & RW & 4,527 \\
20 & ID & 35,599   & 70  & RU & 4,481 \\
21 & IT & 35,050   & 71  & CL & 4,458 \\
22 & JM & 32,835   & 72  & BH & 4,207 \\
23 & AE & 31,700   & 73  & CZ & 4,093 \\
24 & MX & 28,142   & 74  & KW & 3,899 \\
25 & BR & 27,474   & 75  & ET & 3,323 \\
26 & TH & 24,741   & 76  & AG & 3,306 \\
27 & JP & 23,884   & 77  & PA & 3,183 \\
28 & BE & 22,170   & 78  & KH & 3,161 \\
29 & SG & 20,847   & 79  & CR & 3,053 \\
30 & BW & 18,476   & 80  & DO & 2,862 \\
31 & CN & 18,067   & 81  & RO & 2,834 \\
32 & SA & 17,756   & 82  & LS & 2,834 \\
33 & LK & 17,521   & 83  & HU & 2,690 \\
34 & CH & 16,993   & 84  & RS & 2,627 \\
35 & SE & 15,232   & 85  & EC & 2,609 \\
36 & TT & 14,919   & 86  & PE & 2,543 \\
37 & ZW & 13,726   & 87  & VE & 2,369 \\
38 & IL & 12,743   & 88  & MT & 2,367 \\
39 & HK & 12,610   & 89  & IQ & 2,334 \\
40 & NP & 11,049   & 90  & BM & 2,301 \\
41 & QA & 10,766   & 91  & GM & 2,260 \\
42 & TR & 10,591   & 92  & CM & 2,246 \\
43 & PT & 9,851    & 93  & LC & 2,214 \\
44 & BS & 9,324    & 94  & UA & 2,171 \\
45 & DK & 9,121    & 95  & HR & 2,097 \\
46 & KR & 9,117    & 96  & SO & 2,062 \\
47 & BD & 8,998    & 97  & UY & 2,013 \\
48 & CO & 8,994    & 98  & JO & 1,884 \\
49 & TW & 8,956    & 99  & HN & 1,824 \\
50 & ZM & 8,797    & 100 & SZ & 1,818 \\
\hline
\multicolumn{6}{p{8cm}}{other countries with at least 500 tweets (ordered by their frequency): IS, GI, MU, MA, IR, KY, GT, SL, CU, XK, GE, BG, TN, PG, BN, SN, VC, LV, CD, KZ, LU, BZ, PY, SI, SV, VI, AL, GD, SK, AF, LT, MZ, BT, LR, SD, EE, NI, CI, AW, TC, GY, AZ, GU, DZ, BA, MN, KN}\\
\hline
\end{tabular}
$^\#$\url{https://en.wikipedia.org/wiki/ISO_3166-1_alpha-2}
\end{table}

We report the following proportion for different tweet types in the dataset (as shown in Figure \ref{tweet-types}): 14.7\% are original tweets, 66.9\% are retweets, 11.9\% are reply tweets, and 6.5\% are quote tweets. 12.2 million tweets, i.e., approximately 0.87\%, in the dataset have valid geo objects. However, since retweets have NULL geo objects and when therefore excluded, 2.62\% of the (original, reply, and quote) tweets in \textit{BillionCOV} are geotagged.

\begin{table}[ht!]
\centering
\caption{Top 30 sources of tweets.}
\label{sources-list}
\begin{tabular}{|c|>{\centering}m{4cm}|c|}
\hline
\textbf{SN} & \textbf{Source}  & \textbf{Tweets}    \\
\hline
1   & Twitter for iPhone   & 510,113,303 \\
2   & Twitter for Android  & 447,517,798 \\
3   & Twitter Web App      & 300,622,532 \\
4   & Twitter for iPad     & 65,400,331  \\
5   & TweetDeck            & 13,800,728  \\
6   & WordPress.com        & 6,454,537   \\
7   & Hootsuite Inc.       & 5,981,518   \\
8   & dlvr.it              & 5,386,122   \\
9   & IFTTT                & 3,917,407   \\
10  & Twitter Web Client   & 3,214,067   \\
11  & Twitter              & 3,078,620   \\
12  & Buffer               & 2,534,353   \\
13  & Tweetbot for iOS     & 2,475,500   \\
14  & SocialFlow           & 2,431,312   \\
15  & SocialNewsDesk       & 1,905,233   \\
16  & Sprout Social        & 1,389,834   \\
17  & Instagram            & 1,261,162   \\
18  & Paper.li             & 830,365    \\
19  & Echobox              & 821,625    \\
20  & Twitterrific for iOS & 808,327   \\
21 &	Twitter for Mac&	686,768 \\
22 &	TweetCaster for Android&	665,096 \\
23&	Echofon&	631,234 \\
24&	Corona Virus Update Bot&	568,458 \\
25&	True Anthem&	488,188 \\
26&	Tweetbot for Mac&	448,133 \\
27&	LinkedIn&	439,384 \\
28&	Cheap Bots, Done Quick!&	400,776 \\
29&	HubSpot&	391,969 \\
30&	Twitter Media Studio&	390,323 \\
\hline
\multicolumn{3}{p{8cm}}{other sources with at least 200k tweets (ordered by their frequency): Revive Social App, Microsoft Power Platform, FS Poster, Dynamic Signal, Zapier.com, Talon Android, Sprinklr, Blog2Social APP, BLOX CMS, Flamingo for Android, Twibble.io, Corona Updates EA, Tweetlogix, The Social Jukebox, Fenix 2, Salesforce Social Studio, Sendible}\\
\hline
\end{tabular}
\end{table}

Regarding account types (shown in Figure \ref{verified-stats}), 45 million tweets were generated by verified accounts and 1.365 billion tweets by unverified accounts. Verified accounts on Twitter indicate active, notable, and authentic accounts of public interest. We also explored the \textit{source} tweet object for all tweets in \textit{BillionCOV}. Twitter's native apps for iPhone, Android, Web, and iPad are the top 4 sources contributing more than 1.32 billion tweets. The list continues with TweetDeck, WordPress.com, Hootsuite Inc., dlvr.it, IFTTT, and Twitter Web Client as the top 10 sources. The top sources contributing at least 200k tweets are listed in Table \ref{sources-list}.

\subsection{Use cases}
Below we discuss some potential use cases of \textit{BillionCOV}.

\begin{itemize}
    \item The globe or region-specific Twitterverse can be explored through topic modeling \cite{abd2020top} and opinion mining \cite{boon2020public} to examine the public perception of the COVID-19 pandemic and discover the trends and themes of concerns tweeted. The number of topics during a pandemic can be in the thousands, as district-, county-, city-, state-, and country-level large and small events accumulate over time. Exploring evolving nature of events at a large scale can be beneficial to obtain a historical comprehensive situational view of the pandemic in a region. Studying such conversation dynamics also aids in designing future automated information systems for pandemic management.
    \item Twitter users interact through replies, retweets, quotes, likes, followings, etc., forming associations that emerge into complex social network structures, eventually unlocking possibilities for social network analysis \cite{ahmed2020covid,park2020conversations,gruzd2020going,lamsal2021design}. Analysis of such complex networks, at a large scale, assists in investigating multiple network structures within an extensive network~---~such as isolate and broadcast groups~---~and identifying key users and their roles in the network during the pandemic. Some other applications include studying the flow of information surrounding the pandemic, examining if COVID-19 conspiracy theories and propaganda are the results of coordination amongst users or bots, and mining [region$\rightarrow$mention] and [region$\rightarrow$hashtag] relationships for identifying region-specific concerns.
    \item Tweets are composed differently compared to texts from Wikipedia and news articles. Tweets are written concisely, with informal grammar and irregular vocabulary alongside internet abbreviations and hashtags. Training language models on tweet data have produced state-of-the-art results in tweet natural language processing tasks of parts-of-speech tagging, named-entity recognition, and text classification \cite{nguyen2020bertweet}. Original tweets, quote tweets, and reply tweets in \textit{BillionCOV} can be used for training large-scale language models primarily targeted for COVID-19 tweets-related downstream tasks \cite{muller2020covid,nguyen2020bertweet}. Besides, \textit{BillionCOV} can be explored for relevant and informative tweets to generate datasets for downstream applications, especially for supervised learning settings.
    \item Geotagged Twitter conversations have been reported to have variables that Granger-cause the daily COVID-19 confirmed cases time series \cite{lamsal2022twitter}. Latent variables search within geotagged tweets in \textit{BillionCOV} can assist in designing COVID-19 (confirmed/death) cases' forecasting models. Furthermore, \textit{BillionCOV} also has applications in correlation analysis~---~e.g., correlations of COVID-19 (confirmed/death) cases with negative sentiments or sentiment-involved topic-level discussions. Developing methodologies generalizable to future epidemics and pandemics require a large-scale dataset that comprehensively covers a pandemic's conversational dynamics.
\end{itemize}

\subsection{Additional Information}

\textbf{Distribution.} \textit{BillionCOV} is publicly available as an open-access dataset from \textit{IEEE DataPort} at this URL: \texttt{https://dx.doi.org/10.21227/871g-yp65}. A free \textit{IEEE} account is sufficient to download the dataset files.\\

\noindent \textbf{Archival comments.} It is recommended to archive Twitter data by storing the original API responses, i.e. hydrated JSONL data. Tweet objects per requirement can then be later extracted and exported to convenient formats such as CSV for analysis. Programming languages that represent integers with fewer than 64 bits, e.g., Javascript, and integer representation in spreadsheets such as Microsoft Excel, mistranslate the tweet identifiers (64-bit unsigned integers). As a result, the tweet identifiers appear to end with a series of zeroes. Therefore, tweet identifiers should always be loaded and exported as strings instead of integers to avoid generating invalid tweet identifiers. \\

\noindent \textbf{Ethics Statement.} 
While adhering to Twitter's data re-distribution policy, we share tweet identifiers that need to be hydrated to re-create a part or complete dataset locally. The additional data, such as \texttt{is\_reply}, \texttt{is\_retweet}, \texttt{is\_quote}, and \texttt{is\_author\_verified}, provides boolean values of TRUE or FALSE while concealing raw tweet information. Providing the \texttt{country} information for geotagged tweets is equivalent to releasing numerous country-specific tweet collections.\\

\noindent \textbf{Disclaimers.} \cite{lamsalIEEEa} uses Twitter's filtered stream endpoint whose payload returns 1\% of entire Twitter data at a particular time. Also, the number of tweets after hydration can vary as deleted and protected tweets are not retrievable. This dataset should be used only for non-commercial purposes while also strictly adhering to Twitter's policy.

\section{Conclusion}
In this paper, we introduced \textit{BillionCOV}, an enriched global billion-scale English-language COVID-19 tweets dataset, which facilitates researchers to filter tweet identifiers before hydration. \textit{BillionCOV} solves the \textit{proportion} and \textit{redundancy} issues associated with the existing large-scale COVID-19 tweets datasets. We discussed the dataset's curation method and efficient ways to hydrate the tweet identifiers for re-creating the dataset locally. Next, we briefly explored the dataset and discussed its potential use cases. We anticipate that the dataset of this scale with global scope and extended temporal coverage will aid in obtaining a thorough understanding of the pandemic's conversational dynamics.

\section*{Acknowledgements}
This study was supported by the \textit{Melbourne Research Scholarship} from the University of Melbourne, Australia. We are grateful to the \textit{Nectar Research Cloud} for providing us with a large compute instance (80 VCPUs, 720 GB memory, 20TB volume). We are also thankful to \textit{DigitalOcean} for funding the infrastructure needed to maintain \textit{COV19Tweets}\footnote{https://ieee-dataport.org/open-access/coronavirus-covid-19-tweets-dataset} since its inception. We also thank \textit{IEEE DataPort} for supporting \textit{BillionCOV} to be available as an open-access dataset.

\bibliography{aaai22.bib}

\newpage
\appendix

\begin{landscape}
\section{Appendix A: Tweet data dictionary}
\begin{figure}[h!]
\vspace{2cm}
        \includegraphics[width=22cm]{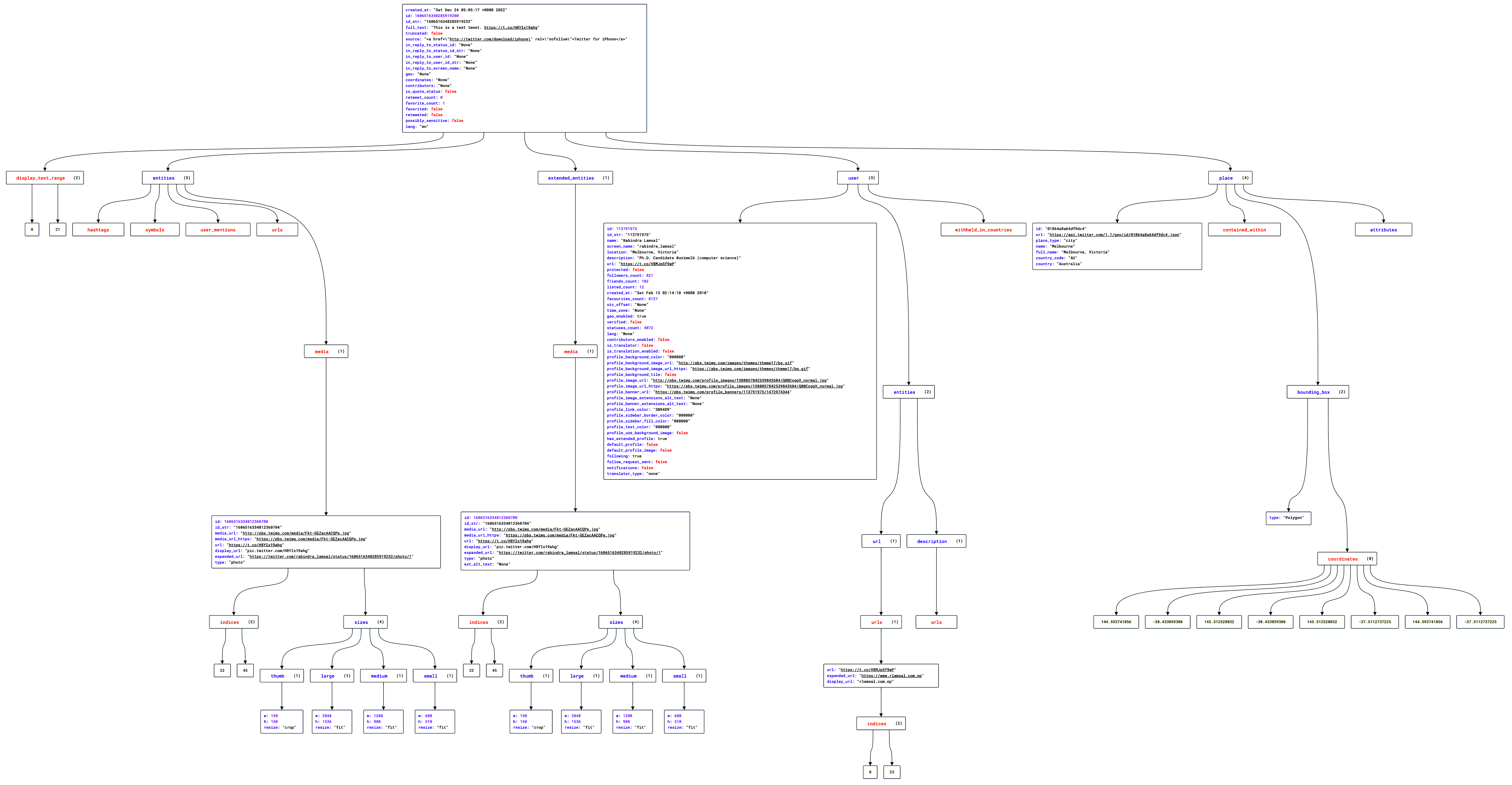}
    \caption{A \textit{tweet data dictionary} generated by \textit{twarc} for the tweet with identifier $1606516340285919232$. The tweet is available at this HTTPS URL: \texttt{{https://twitter.com/rabindra\_lamsal/status/1606516340285919232}}. Note that a tweet can be previewed on a web browser using its identifier: \texttt{https://twitter.com/check/status/identifier\_goes\_here}.} 
\end{figure}
\end{landscape}

\end{document}